\documentclass[aps,pra,showpacs,twocolumn,superscriptaddress,nofootinbib]{revtex4}

\usepackage{graphicx}
\usepackage[normal]{subfigure}
\usepackage{caption}
\usepackage{latexsym}
\usepackage{amsmath}
\usepackage{amssymb}
\usepackage{amsfonts}
\usepackage{color}

\usepackage[utf8]{inputenc}
\usepackage[english]{babel}
\begin{document}

\newcommand{\ket}[1]{\vert #1 \rangle}
\newcommand{\bra} [1] {\langle #1 \vert}
\newcommand{\braket}[2]{\langle #1 | #2 \rangle}
\newcommand{\proj}[1]{\ket{#1}\bra{#1}}
\newcommand{\mean}[1]{\langle #1 \rangle}
\newcommand{\opnorm}[1]{|\!|\!|#1|\!|\!|_2}
\newtheorem{theo}{Theorem}
\newtheorem{lem}{Lemma}
\newtheorem{defin}{Definition}
\newtheorem{corollary}{Corollary}

\title{Majorization preservation of Gaussian bosonic channels}

\author{Michael G. Jabbour}
\author{Ra\'{u}l Garc\'{\i}a-Patr\'{o}n}
\author{Nicolas J. Cerf}
\affiliation{Quantum Information and Communication, Ecole Polytechnique de Bruxelles, CP 165, Universit\'e libre de Bruxelles,
1050 Bruxelles, Belgium}

\begin{abstract}
It is shown that phase-insensitive Gaussian bosonic channels are majorization-preserving over the set of passive states of the harmonic oscillator. This means that comparable passive states under majorization are transformed into equally comparable passive states. The proof relies on a new preorder relation called Fock-majorization, which coincides with regular majorization for passive states but also induces a mean photon number order, thereby connecting the concepts of energy and disorder of a quantum state. As an application, the consequences of majorization preservation are investigated in the context of the broadcast communication capacity of bosonic Gaussian channels. Most of our results being independent of the bosonic nature of the system under investigation, they could be generalized to other quantum systems and Hamiltonians, providing a general tool that could prove useful in quantum information theory and quantum thermodynamics.
\end{abstract}

\pacs{03.67.-a, 03.67.Bg, 42.50.-p, 89.70.-a}

\maketitle

\section{Introduction}

Majorization theory has long been known to play a prominent role in quantum information theory \cite{Nielsen,NielsenVidal}. When a quantum state $\rho$ majorizes another quantum state $\sigma$, denoted as $\rho\succ\sigma$, it means that $\rho$ can be transformed into $\sigma$ by applying a convex combination of unitary operations, that is $\sigma=\sum_i w_i U_i \rho U_i^\dagger$, with $U_i$ being unitaries, $w_i\ge 0$, and $\sum_i w_i=1$. Thus, $\rho\succ\sigma$ means that $\sigma$ is more disordered than $\rho$, and it implies in particular that $S(\sigma)\ge S(\rho)$ where $S$ stands for the von Neumann entropy (more generally, it implies a similar inequality for any concave function of $\rho$). Interestingly, majorization theory also provides a necessary and sufficient condition for the interconversion between pure bipartite states using deterministic Local Operations and Classical Communication (LOCC) \cite{Nielsen,NielsenVidal}. A bipartite pure state $\ket{\psi}$ can be transformed into $\ket{\psi'}$ via a deterministic LOCC if and only if $\rho'\succ\rho$, where $\rho (\rho')$ is the reduced state obtained from its purification $\ket{\psi} (\ket{\psi'})$ by tracing over either one of its two parts. Still another application of majorization is related to separability \cite{NielsenKempe2001}: a separable state $\rho_{AB}$ necessarily obeys $\rho_A \succ \rho_{AB}$ and $\rho_B \succ \rho_{AB}$, which in turn provides a sufficient condition for entanglement that is strictly stronger than the violation of the corresponding entropic conditions $S(\rho_{AB})\ge S(\rho_{A})$ and $S(\rho_{AB})\ge S(\rho_{B})$ \cite{horo96,cerf97}. This result may even be extended to a distillability criterion by noting that  any non-distillable (but possibly bound-entangled) state satisfies the same majorization conditions \cite{hiroshima}.

The importance of majorization theory in continuous-variable quantum information theory was first suggested in \cite{Saikat2008}, specifically in the context of Gaussian bosonic channels. These channels (defined in Sec. II) are ubiquitous in quantum communication theory as they model most optical communication links, such as optical fibers. In \cite{Saikat2008}, Guha was concerned with the classical capacity of these channels (see \cite{Holevo2000}), which was known to require proving a Gaussian minimum entropy conjecture \cite{Giovannetti2004} (now proven in \cite{Giovannetti2014}). Denoting an arbitrary Gaussian bosonic channel by $\Phi[.]$, the conjecture is that $S(\Phi[\ket{\psi}]) \ge S(\Phi[\ket{0}])$ for any input pure state $\ket{\psi}$, where $\ket{0}$ is the vacuum state. The intuition was that a majorization relation $\Phi[\ket{0}] \succ \Phi[\ket{\psi}]$ might be responsible for the conjectured entropic inequality.

The existence of majorization relations in Gaussian bosonic channels was first proven in \cite{Raul2012}, where the quantum-limited amplifier ${\cal A}[.]$  (defined in Sec. II) was proven to obey an infinite ladder of majorization relations when the input state is an individual Fock state, namely  ${\cal A}[\ket{k}] \succ {\cal A}[\ket{k+1}]$, $\forall k\ge 0$. A parametric majorization relations was also proven for varying gain $G$, namely ${\cal A}_G[\ket{k}] \succ {\cal A}_{G+\delta G}[\ket{k}]$ if $\delta G\ge 0$. Then, in \cite{Christos2013}, a similar ladder of majorization relations ${\cal L}[\ket{k}] \succ {\cal L}[\ket{k+1}]$ was shown to hold for a pure loss channel ${\cal L}$ (defined in Sec. II). Later on, in \cite{Mari2014}, the general majorization relation $\Phi[\ket{0}] \succ \Phi[\ket{\psi}]$ was proven  for any input state $\ket{\psi}$ and Gaussian bosonic channel $\Phi$, which generalizes (and implies) the proof \cite{Giovannetti2014} of the above Gaussian minimum entropy conjecture. Finally, the interconversion between pure Gaussian states was investigated based on majorization theory \cite{Giedke2003,Michael2015}, which revealed the existence of surprising situations where a non-Gaussian LOCC is required although the states considered are Gaussian.

In this paper, we first introduce the notion of Fock-majorization, denoted as $ \succ_\mathrm{F}$, which induces a novel (pre)order relation between states of a bosonic mode. We show that Fock-majorization has two powerful properties. Firstly, it induces an order in terms of the  mean energy of the states. Secondly, it coincides with regular majorization for passive states, namely the lowest energy states among isospectral states \cite{passive1,passive2}. Since we focus on the Hamiltonian of an harmonic oscillator, $H=1/2+\hat a^\dagger \hat a$, whose eigenstates are the Fock states, all passive states of a bosonic mode are obviously Fock-diagonal states with decreasing eigenvalues for increasing photon number.

Equipped with this tool, we can prove a new type of intrinsic majorization property in Gaussian bosonic channels, namely the conservation across any channel $\Phi$ of a Fock-majorization relation between any two comparable Fock-diagonal states, that is, $\rho \succ_\mathrm{F} \sigma \Rightarrow \Phi[\rho] \succ_\mathrm{F} \Phi[\sigma] $. This implies in turn that Gaussian bosonic channels preserve regular majorization over the set of passive states of the harmonic oscillator, that is, $\rho \succ \sigma \Rightarrow \Phi[\rho] \succ \Phi[\sigma] $. Finally, we discuss the connection of this result with an open problem related to the broadcast communication capacity of bosonic channels \cite{Saikat2007}.

\section{Gaussian bosonic channels}

An arbitrary Gaussian bosonic channel, denoted as  $\Phi[.]$, is such that if $\rho$ is a Gaussian state, then $\Phi[\rho]$ is a Gaussian state too.
In this paper, we restrict to the class of single-mode phase-insensitive Gaussian bosonic channels, in which the two quadrature components ($\hat x$ and $\hat p$) of the mode operator $\hat a=(\hat x+i \hat p)/\sqrt{2}$  are identically transformed under $\Phi$ in the Heisenberg picture.
A simple example of such a channel is a beam splitter of transmittance $\eta$, which couples the input mode with an environment mode in a vacuum state,
\begin{equation}
\hat a_\mathrm{in} \to \hat a_\mathrm{out } = \sqrt{\eta} \, \hat a_\mathrm{in} + \sqrt{1-\eta} \, \hat a_\mathrm{env}
\end{equation}
where  $\hat a_\mathrm{in}$, $\hat a_\mathrm{env}$, and  $\hat a_\mathrm{out }$ are the bosonic annihilation operators for the input, environment, and output mode, respectively. This is the so-called pure loss channel ${\cal L}_{\eta}$ of transmittance $\eta$, where the Gaussian noise originates from the vacuum fluctuations of the  em field in the environment mode. Another basic phase-insensitive channel is a parametric optical amplifier of gain $G$, which couples the input (or signal) mode with an environment (or idler) mode in a vacuum state according to
\begin{equation}
\hat a_\mathrm{in} \to \hat a_\mathrm{out } = \sqrt{G} \, \hat a_\mathrm{in} + \sqrt{G-1} \, \hat a^\dagger_\mathrm{env}
\end{equation}
In this so-called quantum-limited amplifier channel ${\cal A}_G$ of gain $G$, some Gaussian thermal noise unavoidably affects the output state because of parametric down-conversion. Now if the environment mode is in a thermal state for both cases of a beam-splitter or parametric down-converter, there is an additional Gaussian noise superimposed onto the attenuated or amplified output state, giving rise to a noisy version of channels ${\cal L}_{\eta}$ and ${\cal A}_G$. More generally, any single-mode phase-insensitive  Gaussian bosonic channel $\Phi$ may be decomposed as a suitable sequence of channels ${\cal L}_{\eta}$ and ${\cal A}_G$ \cite{Caruso,Raul2012}.

\section{Fock majorization}

Let us recall that the usual majorization relation is that $\rho \succ \sigma$ if and only if
\begin{equation}
\sum_{i=0}^n \lambda_i^\downarrow \ge \sum_{i=0}^n \mu_i^\downarrow, \;  \forall n\ge 0,
\label{def_majoriz}
\end{equation}
where  $\lambda_i^\downarrow$ and $\mu_i^\downarrow$ are the eigenvalues of $\rho$ and $\sigma$, respectively, that have been sorted by decreasing order. Here, we define that states $\rho$ and $\sigma$ satisfy the Fock-majorization relation $\rho\succ_\mathrm{F} \sigma$  if and only if
\begin{equation}
\mathrm{Tr}(P_n \rho) \ge \mathrm{Tr}(P_n \sigma), \; \forall n\ge 0,
\label{def_Fock_major}
\end{equation}
with $P_n=\sum_{i=0}^n \proj{i}$, which yields a distinct (pre)order relation in state space. In contrast with the definition of regular majorization, the diagonal elements (or eigenvalues if the state is Fock-diagonal) are not ordered by decreasing values but instead by increasing photon number.
Note also that any two Fock states $\ket{n}$ and $\ket{m}$  satisfy the Fock-majorization relation $\proj{n}\succ_\mathrm{F}\proj{m}$ when $n \leq m$, while they are simply equivalent (isospectral) in terms of usual majorization. Another property of Fock majorization is that if $\rho\succ_\mathrm{F} \sigma$ and $\sigma\succ_\mathrm{F} \rho$ are both true, then ${\rm diag}(\rho)={\rm diag}(\sigma)$. By comparison, for regular majorization, if $\rho\succ \sigma$ and $\sigma\succ \rho$, then the states are equivalent (isospectral).

Furthermore, Fock majorization implies an energy relation between comparable states, namely
\begin{equation}
\rho \succ_\mathrm{F} \sigma  \Rightarrow \mathrm{Tr}(\rho \, \hat n) \le \mathrm{Tr}(\sigma \, \hat n),
\label{eq_Fock_major_energy}
\end{equation}
where $\hat n = \hat{a}^{\dagger} \hat{a}$ is the number operator. 
In what follows, we restrict to the case of Fock-diagonal states (especially the passive states of the harmonic oscillator), so we only give the proof of Eq. (\ref{eq_Fock_major_energy}) for these states. Take two Fock-diagonal states
\begin{equation}
\rho=\sum_{i=0}^{N}r_i \proj{i}, \qquad \sigma=\sum_{i=0}^{N}s_i \proj{i},
\end{equation}
whose support is the space spanned by $\{ \ket{0},\cdots \ket{N}\}$. (If their support have unequal sizes, we take the largest size for $N$.)
We assume that $\rho \succ_\mathrm{F} \sigma $, that is
\begin{equation}
\sum_{i=0}^{n} r_i \ge \sum_{i=0}^{n} s_i, \qquad \forall n : 0\le n\le N,
\end{equation}
or equivalently
\begin{equation}
\sum_{i=n}^{N} r_i \le \sum_{i=n}^{N} s_i, \qquad \forall n : 0\le n\le N.
\end{equation}
Summing this expression over $n$ gives
\begin{equation}
\sum_{n=1}^N \sum_{i=n}^{N} r_i \le \sum_{n=1}^N \sum_{i=n}^{N} s_i,
\end{equation}
or, interchanging the two summations,
\begin{equation}
\sum_{i=1}^N \sum_{n=1}^{i} r_i  \le \sum_{i=1}^N \sum_{n=1}^{i} s_i  \Longleftrightarrow  \sum_{i=1}^N i \, r_i \le \sum_{i=1}^N i \, s_i.
\end{equation}
By taking the limit where $N$ tends to infinity, we conclude that the mean energy of $\rho$ is lower than that of $\sigma$, which proves 
Eq. (\ref{eq_Fock_major_energy}). Note that the converse of Eq. (\ref{eq_Fock_major_energy}) is not true.

Finally, it is straightforward to see that Fock majorization $\rho\succ_\mathrm{F} \sigma$ coincides with regular majorization $\rho\succ \sigma$ over the set of passive states. By definition, passive states are diagonal in the energy eigenbasis of the harmonic oscillator (i.e., Fock basis for a bosonic mode) and their eigenvalues are non-increasing with respect to energy, that is, we have
\begin{equation}
\rho=\sum_{i=0}^\infty  r_i \proj{i}, \; {\rm where} \; r_i\ge r_{i+1} , \; \forall i\geq 0,
\end{equation}
for a passive state $\rho$.
Hence,  when restricting to passive states, the Fock-majorization condition Eq.~(\ref{def_Fock_major})
becomes equivalent to the regular majorization relation Eq.~(\ref{def_majoriz}). Otherwise, $\rho\succ_\mathrm{F} \sigma$ and $\rho\succ \sigma$ are unrelated relations.

\section{Passive-preserving channels}

First, a channel $\Phi$ is called Fock-preserving when it is such that if $\rho$ is a Fock-diagonal state, then $\Phi[\rho]$ is also a Fock-diagonal state.
Phase-insensitive Gaussian bosonic channels are well known to be Fock-preserving channels since they map Fock states into mixtures of Fock states \cite{Solomon2011}. A stronger constraint would be to require the channel $\Phi$ to map passive states into passive states, i.e., to be passive-preserving. In order to prove that phase-insensitive Gaussian bosonic channels are indeed passive-preserving\footnote{This property is also proven in \cite{new}, where it is called Fock-preserving, but we give an independent simple proof here.}, we introduce two lemmas. 


\begin{lem} The pure loss channel ${\cal L}$ of arbitrary transmittance exhibits a ladder of Fock-majorization relations
\begin {equation}
{\cal L}\big[ \proj{k} \big]  \succ_\mathrm{F} {\cal L}\big[ \proj{k+1} \big] , \quad  \forall k\ge 0.
\end{equation}
\label{lem_loss}
\end{lem}

It is known that a similar relation holds when replacing Fock-majorization with majorization \cite{Christos2013}. Here, we will adapt this proof in order to derive a Fock-majorization relation. We have
\begin{equation}
\rho^{(k)} \equiv {\cal L}\big[ \proj{k} \big] = \sum_{n=0}^k r_n^{(k)} \proj{n}
\end{equation}
where
\begin{equation}
 r_n^{(k)} = {k \choose n} \eta^n (1-\eta)^{k-n}
\end{equation}
and $\eta$ is the transmittance of channel ${\cal L}$. Majorization was proven in \cite{Christos2013} based on the recurrence relation
\begin{equation}
 r_n^{(k+1)} = \eta \, r_{n-1}^{(k)} + (1-\eta) \, r_{n}^{(k)} , \qquad \forall k\ge0, \forall n\ge 0,
\end{equation}
where the first term in the r.h.s. is taken equal to zero for $n=0$. We can rewrite it as
\begin{equation}
r_{n}^{(k)} -  r_n^{(k+1)} = \eta \left( r_{n}^{(k)} - r_{n-1}^{(k)} \right),
\end{equation}
Hence,
\begin{equation}
\sum_{i=0}^n r_{i}^{(k)} - \sum_{i=0}^n r_i^{(k+1)} = \eta \,  r_{n}^{(k)}  \ge 0
\end{equation}
which gives the Fock-majorization relation $\rho^{(k)}  \succ_\mathrm{F} \rho^{(k+1)}$
in addition to the majorization relation $\rho^{(k)}  \succ \rho^{(k+1)}$ of ref. \cite{Christos2013}.


\begin{lem} The quantum-limited amplifier ${\cal A}$ of arbitrary gain exhibits a ladder of Fock-majorization relations
\begin {equation}
{\cal A}\big[ \proj{k} \big]  \succ_\mathrm{F} {\cal A}\big[ \proj{k+1} \big] , \quad  \forall k\ge 0.
\end{equation}
\label{lem_amp}
\end{lem}

We also use the related majorization property for an amplifier as proven in ref. \cite{Raul2012}. We have
\begin{equation}
\sigma^{(k)} \equiv {\cal A}\big[ \proj{k} \big] = \sum_{n=0}^\infty s_n^{(k)} \proj{n+k},
\end{equation}
where
\begin{equation}
 s_n^{(k)} = {n+k \choose n} t^n (1-t)^{k+1},
\end{equation}
and $t=\tanh^2(r)$ is related to the gain $G=1/(1-t)$ of amplifier $ {\cal A}$, with $r$ being the squeezing parameter.
Majorization was proven in \cite{Raul2012} by using the recurrence relation
\begin{equation}
 s_n^{(k+1)} = t \, s_{n-1}^{(k+1)} + (1-t) \, s_{n}^{(k)} , \qquad \forall k\ge0, \forall n\ge 0,
\end{equation}
where the first term in the r.h.s. is taken equal to zero for $n=0$.
We can rewrite it as
\begin{equation}
s_{n}^{(k)} -  s_n^{(k+1)} = (G-1) \left( s_{n}^{(k+1)} - s_{n-1}^{(k+1)} \right).
\end{equation}
Hence,
\begin{equation}
\sum_{i=0}^n s_{i}^{(k)} - \sum_{i=0}^n s_i^{(k+1)} = (G-1) \,  s_{n}^{(k+1)}  \ge 0
\end{equation}
giving the Fock-majorization relation $\sigma^{(k)}  \succ_\mathrm{F} \sigma^{(k+1)}$
in addition to the majorization relation $\sigma^{(k)}  \succ \sigma^{(k+1)}$ \cite{Raul2012}.

\bigskip
The following theorem is then a key to determine whether some channel is passive-preserving. 

\begin{theo} A channel $\Phi$ is passive-preserving if and only if its dual channel $\Phi^\dagger$
obeys the ladder of Fock-majorization relations
\begin {equation}
\Phi^\dagger \big[ \proj{k} \big]  \succ_\mathrm{F} \Phi^\dagger \big[ \proj{k+1} \big] , \quad  \forall k\ge 0.
\label{eq-Fock-maj-ladder}
\end{equation}
\label{theorem-1}
\end{theo}
Proof. Equation (\ref{eq-Fock-maj-ladder}) is equivalent to
\begin {equation}
\mathrm{Tr} \left( P_n \Phi^\dagger \big[ \proj{k} \big] \right) \ge \mathrm{Tr} \left( P_n \Phi^\dagger \big[ \proj{k+1} \big] \right),  \; \forall n\ge 0,
\label{eqtheodualpass}
\end {equation}
where $P_n=\sum_{i=0}^n \proj{i}$. The dual expression gives
\begin {equation}
\mathrm{Tr} \left(  \proj{k} \Phi [ P_n ] \right) \ge \mathrm{Tr} \left(  \proj{k+1} \Phi [ P_n ] \right).
\label{eq_dual_ineq}
\end {equation}
Now, assume that the input of channel $\Phi$ is a passive state (of the harmonic oscillator)
\begin{equation}
\rho=\sum_{n=0}^{\infty}r_n \proj{n},  \qquad \mathrm{with~} r_n\ge r_{n+1}.
\end{equation}
It can also be rewritten as
\begin{equation}
\rho=\sum_{n=0}^{\infty} e_n P_n,  \qquad \mathrm{with~} e_n=r_n-r_{n+1}.
\end{equation}
where $e_n\ge 0$, $\forall n \ge 0$, since $\rho$ is passive. Then, we may take the convex combination of inequalities (\ref{eq_dual_ineq}) with weights $e_n$ and $n$ going from $0$ to $\infty$, resulting in
\begin {equation}
\mathrm{Tr} \left(  \proj{k} \Phi [ \rho ] \right) \ge \mathrm{Tr} \left(  \proj{k+1} \Phi [ \rho ] \right).
\end {equation}
Hence, the output state $\Phi [ \rho ]$ is passive, so that channel $\Phi$ is indeed passive-preserving.
Conversely, it is trivial to see that $\Phi$ being passive-preserving implies Eq.~(\ref{eq_dual_ineq}) and consequently  Eq.~(\ref{eq-Fock-maj-ladder}). $\Box$

\begin{corollary}
Phase-insensitive Gaussian bosonic channels are passive preserving.
\end{corollary}

Using Lemma \ref{lem_loss} and \ref{lem_amp} together with Theorem \ref{theorem-1} we obtain that the pure loss channel ${\cal L}={\cal A}^\dagger$ and quantum-limited amplifier ${\cal A}={\cal L}^\dagger$ are passive preserving.
Then, the corollary follows from the fact that any phase-insensitive Gaussian bosonic channel $\Phi$ can be expressed as the concatenation of a pure loss channel ${\cal L}$ and a quantum-limited amplifier, i.e., $\Phi = {\cal A} \circ {\cal L}$, and that passive-preservation is transitive over channel composition.

\section{Fock-majorization preserving channels}

A Fock-preserving channel $\Phi$ is called Fock-majorization preserving provided it is such that if $\rho \succ_\mathrm{F}  \sigma$ with $\rho$ and $\sigma$ being Fock-diagonal states, then $\Phi[\rho] \succ_\mathrm{F} \Phi[\sigma]$. In order to prove that phase-insensitive Gaussian bosonic channels are Fock-majorization preserving, we need to prove the following theorem.  


\begin{theo}  A channel $\Phi$ is Fock-majorization preserving if and only if it obeys the ladder of Fock-majorization relations
\begin {equation}
\Phi\big[ \proj{k} \big]  \succ_\mathrm{F} \Phi\big[ \proj{k+1} \big] , \quad  \forall k\ge 0,
\end{equation}
\label{theorem-2}
\end{theo}
Proof. We start with two Fock-diagonal states
\begin{equation}
\rho=\sum_{i=0}^{N}r_i \proj{i}, \qquad \sigma=\sum_{i=0}^{N}s_i \proj{i},
\end{equation}
whose supports is the space spanned by $\{ \ket{0},\cdots \ket{N}\}$. (If their supports have unequal sizes, we take the largest size for $N$.)
We assume that $\rho$ Fock-majorizes $\sigma$, that is
\begin{equation}
\rho \succ_\mathrm{F} \sigma \quad \Longleftrightarrow  \quad R_n \ge S_n, \forall n\ge 0,
\label{eq-theor1-hypothesis}
\end{equation}
where
\begin{equation}
 R_n=\mathrm{Tr}(P_n \rho)=\sum_{i=0}^{n} r_i, \quad S_n= \mathrm{Tr}(P_n \sigma)=\sum_{i=0}^{n} s_i.
\end{equation}
We want to prove that $\Phi[\rho]$ Fock-majorizes $\Phi[\sigma]$, that is,
\begin{equation}
\Phi[\rho] \succ_\mathrm{F} \Phi[\sigma]  \quad \Longleftrightarrow  \quad A_n \ge B_n, \forall n\ge 0
\label{eq-theor1-to-prove}
\end{equation}
where
\begin{equation}
\left\lbrace \begin{aligned}
& A_n=\mathrm{Tr}(P_n \Phi[\rho])=\sum_{i=0}^{n} r_i P_n^{(i)} \\
 & B_n= \mathrm{Tr}(P_n \Phi[\sigma])=\sum_{i=0}^{n} s_i P_n^{(i)}
\end{aligned} \right.
\end{equation}
with
\begin{equation}
P_n^{(i)} = \mathrm{Tr} \left( P_n \Phi\big[\proj{i}\big] \right).
\end{equation}
Now, we define the quantities
\begin{equation}
\alpha_n^{(k)}= R_k P_n^{(k)} +  \sum_{i=k+1}^{N} r_i P_n^{(i)} , \qquad k=0,\cdots N,
\end{equation}
where the second term in the right-hand side is taken equal to zero when $k=N$, so that $\alpha_n^{(N)}=R_N P_n^{(N)}$. Similarly, we define
\begin{equation}
\beta_n^{(k)}= S_k P_n^{(k)} +  \sum_{i=k+1}^{N} s_i P_n^{(i)} , \qquad k=0,\cdots N,
\end{equation}
with the convention  $\beta_n^{(N)}=S_N P_n^{(N)}$.
The Fock-majorization relation we need to prove, Eq.~(\ref{eq-theor1-to-prove}), is equivalent to
\begin{equation}
\alpha_n^{(0)} \ge \beta_n^{(0)}, \qquad  \forall n\ge 0
\end{equation}
corresponding to $k=0$. We will now prove
\begin{equation}
\alpha_n^{(k)} \ge \beta_n^{(k)}, \qquad  \forall n\ge 0
\label{eq_recurrence}
\end{equation}
by recurrence in $k$, starting from $k=N$ and ending at $k=0$. We have trivially
$\alpha_n^{(N)} \ge \beta_n^{(N)}$, $\forall n\ge 0$, since $R_N=S_N=1$. Now, we assume that
\begin{equation}
\alpha_n^{(k+1)} \ge \beta_n^{(k+1)}, \qquad  \forall n\ge 0
\label{eq_start_recurrence}
\end{equation}
which can be rewritten as
\begin{equation} \nonumber
R_{k+1} P_n^{(k+1)} +  \sum_{i=k+2}^{N} r_i P_n^{(i)}  \ge S_{k+1} P_n^{(k+1)} +  \sum_{i=k+2}^{N} s_i P_n^{(i)}.
\end{equation}
Using $R_{k+1}=R_{k}+r_{k+1}$ and $S_{k+1}=S_{k}+s_{k+1}$, we reexpress it as
\begin{equation}
R_{k} P_n^{(k+1)} +  \sum_{i=k+1}^{N} r_i P_n^{(i)}  \ge S_{k} P_n^{(k+1)} +  \sum_{i=k+1}^{N} s_i P_n^{(i)}
\end{equation}
which is equivalent to
\begin{equation}
R_{k} \left( P_n^{(k+1)}- P_n^{(k)} \right) +  \alpha_n^{(k)}  \ge S_{k} \left( P_n^{(k+1)}- P_n^{(k)} \right) +  \beta_n^{(k)}
\end{equation}
or simply
\begin{equation}
 \alpha_n^{(k)}  - \beta_n^{(k)}  \ge  (R_{k} -S_{k}) \left( P_n^{(k)}- P_n^{(k+1)} \right)
 \label{eq-last-alpha-beta}
\end{equation}
Since  $\rho$ Fock-majorizes $\sigma$ by hypothesis [Eq. (\ref{eq-theor1-hypothesis})], we have $R_{k} -S_{k} \ge 0$, $\forall k\ge 0$.
If  $\Phi\big[ \proj{k} \big]$ Fock-majorizes $\Phi\big[ \proj{k+1} \big]$, which means that
$P_n^{(k)} - P_n^{(k+1)} \ge 0 $, $\forall n\ge 0$, then the right-hand side of Eq. (\ref{eq-last-alpha-beta}) is greater than zero.
Thus, Eq. (\ref{eq_start_recurrence}) implies Eq. (\ref{eq_recurrence}), which concludes the recurrence in $k$
and proves Eq. (\ref{eq-theor1-to-prove}). Conversely, it is trivial to see that Fock-majorization preservation for channel $\Phi$ implies the ladder of Fock-majorization relations since individual Fock states  satisfy the Fock-majorization relation $\proj{n} \succ_\mathrm{F} \proj{n+1}$, $\forall n\ge 0$. $\Box$

\begin{corollary} Phase-insensitive Gaussian bosonic channels  are Fock-majorization-preserving.
\end{corollary}

We use again the fact that any phase-insensitive Gaussian bosonic channel $\Phi$ can be expressed as the concatenation $\Phi = {\cal A} \circ {\cal L}$
and that Fock-majorization preservation is transitive over channel composition.

\begin{corollary} Phase-insensitive Gaussian bosonic channels  are majorization-preserving over the set of passive states.
\end{corollary}

As a consequence of the equivalence between Fock-majorization and regular majorization over the set of passive states, a Fock-majorization preserving channel is necessarily also majorization-preserving over the set of passive states provided it is passive-preserving. Since phase-insensitive Gaussian bosonic channels are passive-preserving (Corollary 1) and Fock-majorization preserving (Corollary 2), we conclude that they preserve regular majorization over the set of passive states.


\section{Discussion and conclusion}

We have introduced the notion of Fock-majorization, which induces a novel (pre)order relation between states of the harmonic oscillator and
coincides with regular majorization for passive states, namely the lowest energy states among isospectral states. As a notable application of this tool, we have shown that phase-insensitive Gaussian bosonic channels preserve majorization over the set of passive states. This novel majorization relation nicely complements the one very recently found in \cite{new}. There, it was shown that among all isospectral states $\rho$ at the input of a Gaussian bosonic channel $\Phi$, the passive state $\rho^\downarrow$ produces an output that majorizes all other outputs \footnote{We had independently conjectured this majorization property, without giving a proof.}, namely $\Phi[\rho^\downarrow] \succ \Phi[\rho]$. Instead, we consider here two input states that have different spectra but are both passive, $\rho^\downarrow$ and $\sigma^\downarrow$, and have been able to prove that $\rho^\downarrow \succ \sigma^\downarrow$ implies $\Phi[\rho^\downarrow] \succ \Phi[\sigma^\downarrow]$. This reflects the fact Gaussian bosonic channels exhibit quite a wide range of fundamental majorization properties, going well beyond what was originally expected in ref. \cite{Saikat2008}. Combining these various results may help pave the way to solving some of the open problems in the field of Gaussian bosonic channels.

As a matter of fact, it is worth noting that if Fock-majorization was a full order instead of a preorder relation, our result would solve for instance the broadcast communication capacity of phase-insensitive Gaussian bosonic channels. Indeed, this problem relies on the conjecture that for any input state $\rho$ satisfying the entropy constraint $S(\rho)\geq S$, we have $S \left( \Phi[\rho] \right) \ge S \left( \Phi[\tau] \right)$, where $\tau$ is the (Gaussian) thermal state that has the same entropy $S(\tau)=S$  \cite{Saikat2007}. As a result of ref. \cite{new}, we may restrict ourselves to the minimization over passive states $\rho^\downarrow$ at the input. Then, as a consequence of our Corollary 3 and if Fock-majorization was a full order relation, the passive state at the input that gives the output state with lowest entropy would necessarily have to Fock-majorize all other input states. Next, using the energy order imposed by Fock-majorization, Eq. (\ref{eq_Fock_major_energy}), we would conclude that the optimal input state satisfying the entropy constraint $S \geq S(\rho)$ should also have minimum energy. Since the thermal state $\tau$ has the lowest possible energy for a given input entropy $S$, this would conclude the proof of the previously mentioned conjecture. Unfortunately, Fock-majorization is a preorder relation, which means that there exist pairs of incomparable states that neither satisfy $\rho \succ_\mathrm{F} \sigma$ nor $\sigma \succ_\mathrm{F}  \rho$.
Hence, the previous argument is not conclusive, despite providing further evidence of the conjecture being true. It also suggests that understanding the properties of states that are incomparable under majorization to the thermal state is a crucial step in solving the above conjecture.

Finally, we would like to stress that all proofs in this work, except for lemma 1 and 2, are independent of the specific nature of the system (i.e., the harmonic oscillator Hamiltonian for a bosonic mode). Therefore, we believe that the application of Fock-majorization relations could be extended to other quantum systems and Hamiltonians, providing a novel general tool that could prove very useful in quantum information theory. Furthermore, since Fock majorization induces a mean energy order,  on the one hand, and is connected to regular majorization and the resulting notion of entropy, on the other hand, we anticipate that it will also find interesting applications in the field of quantum thermodynamics.




\section{Acknowledgments}
\acknowledgments
We thank S. Guha and J.H. Shapiro for useful discussions.
This work was supported by the F.R.S.-FNRS under Project No. T.0199.13 and by the Belgian Federal IAP program under Project No. P7/35 Photonics@be. M.G.J. acknowledges support from FRIA foundation. R.G.-P. is Research Associate of the F.R.S.-FNRS.


\begin{thebibliography}{99}

\bibitem{Nielsen} M. A. Nielsen, Phys. Rev. Lett. \textbf{83,} 436 (1999).

\bibitem{NielsenVidal} M. A. Nielsen and G. Vidal, Quant. Inf. Comp. \textbf{1,} 76 (2001).

\bibitem{NielsenKempe2001} M. A. Nielsen and J. Kempe, Phys. Rev. Lett. {\bf 86}, 5184 (2001).

\bibitem{horo96} R. Horodecki, P. Horodecki, and M. Horodecki, Phys. Lett. A \textbf{210}, 377 (1996).

\bibitem{cerf97} N.J. Cerf and C. Adami, Phys. Rev. Lett. \textbf{79},  5194 (1997).

\bibitem{hiroshima} T. Hiroshima, Phys. Rev. Lett. \textbf{91}, 057902 (2003).

\bibitem{Saikat2008} S. Guha, {\it Multiple-user quantum information theory for optical communication channels}, (Ph.D. thesis, Massachusetts Institute of Technology, 2008).

\bibitem{Holevo2000} A. S. Holevo and R. F. Werner, Phys. Rev. A \textbf{63}, 032312 (2001).

\bibitem{Giovannetti2004} V. Giovannetti, S. Guha, S. Lloyd, L. Maccone, and J. H. Shapiro, Phys. Rev. A \textbf{70}, 032315 (2004).

\bibitem{Giovannetti2014} V. Giovannetti, R. Garc\'{\i}a-Patr\'{o}n, N. J. Cerf, and A. S. Holevo, Nature Photonics \textbf {8}, 796 (2014).

\bibitem{Raul2012} R. Garc\'{\i}a-Patr\'{o}n, C. Navarrete-Benlloch, S. Lloyd, J. H. Shapiro and N. J. Cerf, Phys. Rev. Lett. \textbf{108}, 110505 (2012).

\bibitem{Christos2013} C. N. Gagatsos, O. Oreshkov, and N. J. Cerf, Phys. Rev. A \textbf{87}, 042307 (2013).

\bibitem{Mari2014} A. Mari, V. Giovannetti, and A.S. Holevo, Nature Comm. \textbf {5}, 3826 (2014).

\bibitem{Giedke2003} G. Giedke, J. Eisert, J. I. Cirac and M. B. Plenio, Quant. Inf. Comp. \textbf{3,} 211 (2003).

\bibitem{Michael2015} M. G. Jabbour, R. Garcia-Patron, and N. J. Cerf, Phys. Rev. A \textbf{91}, 012316 (2015).

\bibitem{passive1} W. Pusz and S. L. Woronowicz, Commun. Math. Phys. \textbf{58}, 273 (1978).

\bibitem{passive2} A. Lenard, J. Stat. Phys. \textbf{19}, 575 (1978).

\bibitem{Saikat2007} S. Guha, J. H. Shapiro, and B. I. Erkmen, Phys. Rev. A \textbf{76}, 032303 (2007).

\bibitem{Caruso} F. Caruso, V. Giovanetti, and A. S. Holevo, New J. Phys. {\bf 8}, 310 (2006).

\bibitem{Solomon2011} J. Solomon Ivan, Krishna Kumar Sabapathy, and R. Simon,
Phys. Rev. A \textbf{84}, 042311 (2011).

\bibitem{new} G. De Palma, D. Trevisan and V. Giovannetti, arXiv:1511.00293 [quant-ph].

\end{thebibliography}
\end{document}